# Three-dimensional imaging of direct-written photonic structures

Graham D. Marshall,[1,*] Alexander Jesacher,[2] Anisha Thayil,[2] Michael J. Withford,[1] and Martin Booth[2]

[1]*Centre for Ultrahigh bandwidth Devices for Optical Systems, MQ Photonics Research Centre, Department of Physics and Astronomy, Macquarie University, NSW 2109, Australia*
[2]*Department of Engineering Science, University of Oxford, Parks Road, Oxford, OX1 3PJ, UK*
[*]*Corresponding author: graham.d.marshall@gmail.com*

**Abstract**

Third harmonic generation microscopy has been used to analyze the morphology of photonic structures created using the femtosecond laser direct-write technique. Three dimensional waveguide arrays and waveguide-Bragg gratings written in fused-silica and doped phosphate glass were investigated. A sensorless adaptive optical system was used to correct the optical aberrations occurring in the sample and microscope system, which had a lateral resolution of less than 500 nm. This non-destructive testing method creates volume reconstructions of photonic devices and reveals details invisible to other linear microscopy and index profilometry techniques.

**Main text**

The femtosecond laser direct-write technique [1] is an established method for the fabrication of photonic waveguide systems in dielectric media such as single and multi component glasses. Examples of devices fabricated include 3-dimensional waveguide networks [2], couplers [3], Bragg gratings [4], optical amplifiers [5] and lasers [6]. The direct-write process relies on the highly nonlinear interaction between a focused ultrafast laser pulse and the local medium to create a localized refractive index (RI) change, and it is the nonlinear nature of this process that confines the modified region to the focal volume. There are a number of diagnostic tools that can be used to study such laser written waveguides. Transmission differential interference contrast (TDIC) microscopy is a non-destructive transverse (*xy* plane) viewing technique and can provide useful insights into the dimensions of waveguide structures along with some qualitative information on the RI contrast. A more quantitative tool, refractive index profilometry (RIP), can be applied wherein the refractive index contrast of a longitudinal section (*yz* plane) of the waveguide can be measured—however this technique is limited to measurements at polished end facets of waveguide structures and requires the device to be cross-sectioned and therefore destroyed in many cases. In this Letter we present a non-destructive method for the complete volume reconstruction of laser-written photonic structures. We have used third-harmonic generation (THG) microscopy [7] in an adaptive-optical (AO) microscope to study the three dimensional morphology of waveguide arrays and waveguide-Bragg grating (WBG) structures. Harmonic generation microscopy, a technique often used for the study of biological specimens and thin films, relies on the intrinsic non-linearity of a sample or medium interface and is thus label free. This makes it ideally suited to glass samples whose chemistry is unlikely to provide a useful fluorophore that is sensitive to the local refractive index. The non-linear processes that lead to the generation of the third-harmonic signal in the media studied herein are weak and so to improve the microscope's sensitivity and spatial resolution we employed an AO system to correct for the aberrations induced in the sample and imperfections of the elements in the optical train [8]. The strength of the third-harmonic signal obtained from within a photonic device could be used as a metric to optimize the wavefront correction applied to the probe laser beam and the process of probing samples caused no detectable change in the samples' properties or performance.

The femtosecond laser direct-write system used to fabricate the waveguide and WBG samples described herein has been detailed previously [9,10] and consisted of an amplified Ti:Sapphire femtosecond laser producing 120 fs pulses with a wavelength of 800 nm at a repetition rate of 1 kHz. Sub-millijoule pulse energies from this laser were passed through a quarter-wave plate to circularly polarize the laser. The laser beam was then shaped using a slit aperture and focused using a 40× 0.6 NA microscope objective into the glass sample which was held on a three-axis translation stage system. This setup was used to fabricate a stretched hexagonal waveguide array approximately 170 µm below the surface of a 3 mm thick high purity fused silica substrate using the manufacturing conditions described in [10]. The fabrication depth was chosen to coincide with the thickness of cover slip that the fabrication objective was corrected for. To enable a comparison of the THG microscope data with the refractive index profile the sample was sectioned and polished to enable RIP of the waveguide array. In a similarly dimensioned 5% wt. Ytterbium doped phosphate glass sample, WBGs with periods ranging from ~350 nm to 2 µm were fabricated using the methods described in [9]. In addition to the THG analysis, this sample was studied using TDIC microscopy and the strength of several of the WBGs measured in transmission using a swept-wavelength system. The THG microscope system used in this study was similar to that detailed in [8]. The system consisted of a 100 MHz chromium forsterite laser operating at 1235 nm producing 65 fs output pulses; the laser beam was directed to a galvanometer *xy*-scan mirror pair before being reflected by a deformable membrane mirror which was imaged onto the input pupil of the focusing objective. A water immersion objective (40× 1.15 NA) was used as the microscope focusing optic because it had a high transmission efficiency at 1235 nm. The laser power focused into the sample was ~35 mW. Depth (or *z*-axis) scanning was achieved by mounting the sample on a piezo actuated stage. A high numerical aperture (NA=1.4) oil immersion condenser lens collected the THG signal which was detected using a photomultiplier tube after filtering from the fundamental light. The AO correction setup and sample scanning was automated and required approximately 10 minutes to complete a typical volume image stack.

A still frame from the three dimensional rendering of the hexagonal waveguide array acquired by the THG microscope is shown in Fig. 1(a) and the animation may be seen in [Media 1](). By comparing the THG microscope data with the measured RIP shown in Fig. 1(b) it can be seen that the volume rendering clearly resolves the waveguides and faithfully reproduces the 3D arrangement of the array. The high numerical aperture of the microscope allows one to obtain image layers through surrounding regions that contain waveguides without signs of perturbation of the probe beam. The shape of the waveguides in the volume rendering appears slightly stretched in the vertical dimension. This apparent asymmetry is caused by the slightly reduced resolution of the microscope in the axial direction of the probe laser beam; in the setup the axial resolution was approximately 1.3 µm, while the lateral resolution was better than 500 nm. Another feature of the reconstruction in Fig. 1(a) is that the waveguides appear 'hollow' due a reduction of the THG signal obtained at the centre of the waveguide. This may be understood when one considers the conditions required for THG as explained in [7]. Under tight focusing conditions THG is only obtained in regions where the RI or third-order susceptibility ($\chi^{(3)}$) of the material is non-uniform (hence the regions around the waveguides in Fig. 1(a) and Fig. 3(a) are black). As can be seen in Fig. 1(b) the waveguides have an approximately 'flat-top' RI profile and therefore the THG efficiency at the centre of the waveguide is lower than at the edges. In these observations it would be natural to initially assume that the process causing the THG is dominated by the ~2% change in the local refractive index however proportionately larger changes in third-order nonlinear RI ($n_2$) have been observed in glass waveguide structures fabricated using processing conditions similar to ours. In [11] the $n_2$ measured in direct-written fused silica waveguides was observed to *decrease* by 74% of the bulk glass value. Clearly the measured RI change of +0.13% observed in the waveguides detailed in [11] did not lead to the small *increase* in $n_2$ one might predict from empirical scaling laws such as Miller's rule, and our investigations into the relative contributions to THG from gradients in $\chi^{(3)}$ and RI change in laser written structures are ongoing.

So far we have examined photonic structures that have micrometer scale features that are larger than the predicted resolution limit of the THG microscope. To study other laser written devices of interest and determine the resolution limit of the microscope setup we investigated a series of WBGs. A range of gratings with different periods were imaged; these included 1st to 4th order gratings designed for operation at $\lambda_B = 1.54\,\mu m$ and 1.06 µm (wavelengths of common use in monolithic waveguide laser platforms) and the periods of these devices ranged from 347nm to 2.01 µm. The theoretical resolution (based on NA) of the microscope was 670 nm however some improvement in the resolution of the optical system may be expected given that the nonlinear THG process will reduce the effective probe spot size produced by the scanning laser. Single images from the volume stack for each of the WBGs with resolved grating periods are shown in Fig. 2. The microscope was unable to resolve the 347 nm periods of the shortest wavelength 1st order grating however the RI modulations of the 503 nm grating are clearly visible. In the 693 and 503 nm period gratings the regions of greatest THG intensity are located at the edges of the grating where the gradients in RI or $\chi^{(3)}$ are effectively the greatest with respect to the finite sized probe laser spot. A frame from the 3D volume rendering of the 1.01 µm period WBG and a TDIC micrograph of the same device are shown in Fig. 3(a) and (b) respectively. The full animation of the WBG section may be found in [Media 2](). The 3D view of the grating is vertically sectioned at the mid-point to reveal the morphology of the inside of the WBG. The same 'hollowing' phenomenon affects the WBG as it did the waveguide array however this does not detract from other observations. It can be seen that the WBG has a triangular shape in its lateral section; this characteristic is understood to be due to the spherical aberration present in the waveguide fabrication system that results from using a metallurgical (surface) objective to focus the laser rather than one corrected for sub-surface focusing. The grating is also revealed as having planes of RI modulation that are slightly curved and that appear to bifurcate and coalesce a few microns from the top and bottom of the modified region. These characteristics are visible in the 1.01 µm and 693 nm period WBGs and may be due to the presence of one grating plane affecting the writing process for its neighbor. For comparison, a TDIC image of the grating is shown in Fig. 3(b), the location of the image in the axial plane of the THG microscope is indicated in Fig. 3(a). To ascertain whether the process of THG analysis (and in particular the exposure of the WBG sample to additional ultrafast laser radiation) modifies a sample we characterized the grating transmission properties of the $\lambda_B = 1.54\,\mu m$ gratings before and after analysis. In these measurements we did not detect any modification of the gratings' spectrum;

neither were we able to observe (using the TDIC microscope) any bulk RI change caused by the scanning probe laser. These results lead us to believe that the THG method does not cause changes to the samples under study therefore the technique is non-destructive.

We have employed THG microscopy to analyze the 3-dimensional arrangement and morphology of ultrafast laser written photonic devices. This is, to our knowledge, the first application of harmonic generation microscopy to directly-written photonic systems and we have shown that, having developed a system with lateral resolution of less than 500 nm, it is possible to study important waveguides and WBG devices without modification or the requirement to section (and thus destroy) an optical sample. The THG microscope system employed was able to study samples several millimeters thick and had threshold sensitivity sufficient to view waveguides in fused silica with a peak RI change of $9\times10^{-3}$ in a 4.0 µm diameter guide ($1/e^2$ width). Third harmonic generation microscopy has been shown to be a useful tool that will assist the development of complex multidimensional optical waveguide systems.

This work was supported by the Australian Research Council (through their Centres of Excellence program), and by the UK EPSRC (EP/E055818/1) and BBSRC (BB/F011512/1). A.J. was supported by the Austrian Science Fund (J2826-N20).

**References**
1. K. M. Davis, K. Miura, N. Sugimoto, and K. Hirao, "Writing waveguides in glass with a femtosecond laser," Opt Lett **21**(21), 1729-1731 (1996), http://dx.doi.org/10.1364/OL.21.001729.
2. A. Szameit, D. Blomer, J. Burghoff, T. Pertsch, S. Nolte, and A. Tünnermann, "Hexagonal waveguide arrays written with fs-laser pulses," Appl Phys B-Lasers O **82**(4), 507-512 (2006), http://dx.doi.org/10.1007/s00340-005-2127-4.
3. S. M. Eaton, W. Chen, L. Zhang, H. Zhang, R. Iyer, J. S. Aitchison, and P. R. Herman, "Telecom-band directional coupler written with femtosecond fiber laser," Ieee Photonic Tech L **18**(17-20), 2174-2176 (2006), http://dx.doi.org/10.1109/LPT.2006.884241
4. H. B. Zhang, S. M. Eaton, J. Z. Li, and P. R. Herman, "Femtosecond laser direct writing of multiwavelength Bragg grating waveguides in glass," Opt Lett **31**(23), 3495-3497 (2006), http://dx.doi.org/10.1364/OL.31.003495.
5. G. Della Valle, R. Osellame, N. Chiodo, S. Taccheo, G. Cerullo, P. Laporta, A. Killi, U. Morgner, M. Lederer, and D. Kopf, "C-band waveguide amplifier produced by femtosecond laser writing," Opt Express **13**(16), 5976-5982 (2005), http://dx.doi.org/10.1364/OPEX.13.005976.
6. G. D. Marshall, P. Dekker, M. Ams, J. A. Piper, and M. J. Withford, "Directly written monolithic waveguide laser incorporating a distributed feedback waveguide-Bragg grating," Opt Lett **33**(9), 956-958 (2008), http://dx.doi.org/10.1364/OL.33.000956.
7. Y. Barad, H. Eisenberg, M. Horowitz, and Y. Silberberg, "Nonlinear scanning laser microscopy by third harmonic generation," Appl Phys Lett **70**(8), 922-924 (1997), http://dx.doi.org/10.1063/1.118442.
8. A. Jesacher, A. Thayil, K. Grieve, D. Debarre, T. Watanabe, T. Wilson, S. Srinivas, and M. Booth, "Adaptive harmonic generation microscopy of mammalian embryos," Opt Lett **34**(20), 3154-3156 (2009), http://dx.doi.org/10.1364/OL.34.003154.
9. M. Ams, P. Dekker, G. D. Marshall, and M. J. Withford, "Monolithic 100 mW Yb waveguide laser fabricated using the femtosecond-laser direct-write technique," Opt Lett **34**(3), 247-249 (2009), http://dx.doi.org/10.1364/OL.34.000247.
10. G. D. Marshall, A. Politi, J. C. F. Matthews, P. Dekker, M. Ams, M. J. Withford, and J. L. O'Brien, "Laser written waveguide photonic quantum circuits," Opt Express **17**(15), 12546-12554 (2009), http://dx.doi.org/10.1364/OE.17.012546.
11. D. Blomer, A. Szameit, F. Dreisow, T. Schreiber, S. Nolte, and A. Tünnermann, "Nonlinear refractive index of fs-laser-written waveguides in fused silica," Opt Express **14**(6), 2151-2157 (2006), http://dx.doi.org/10.1364/OE.14.002151.

**Figures**

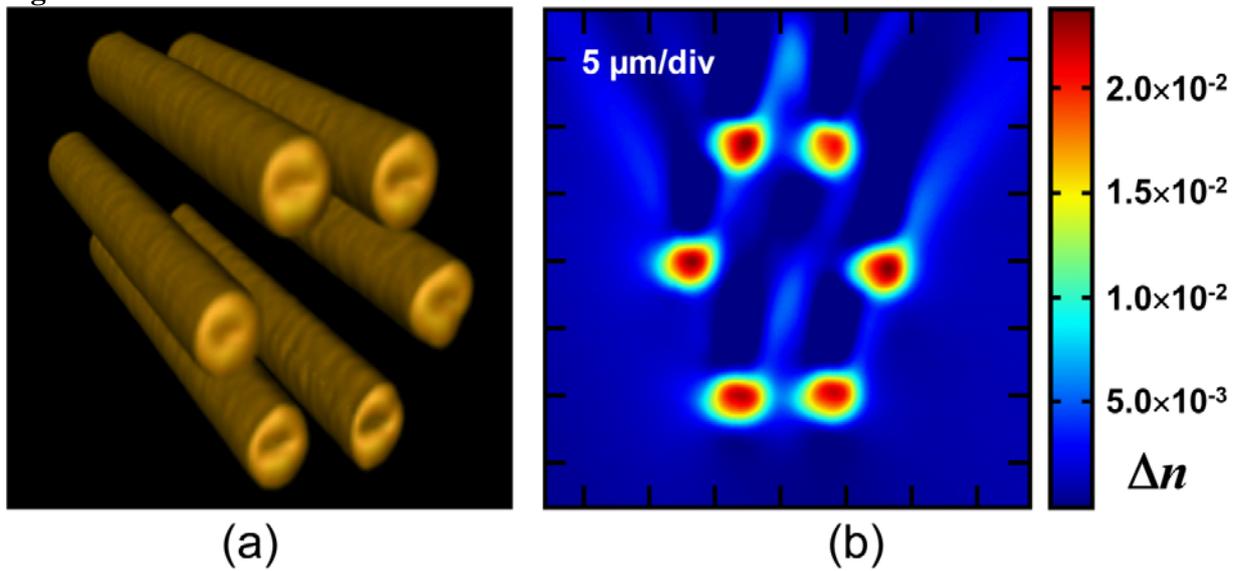

Fig. 1 (a) A still image from the animation obtained using the THG microscope images of the hexagonal waveguide-array. Full animation shown in Media 1 (800 KB). (b) A refractive index profile of the end facet of the array. The vertically arranged 'shadows' of the waveguides are an artifact of the RIP measurement method.

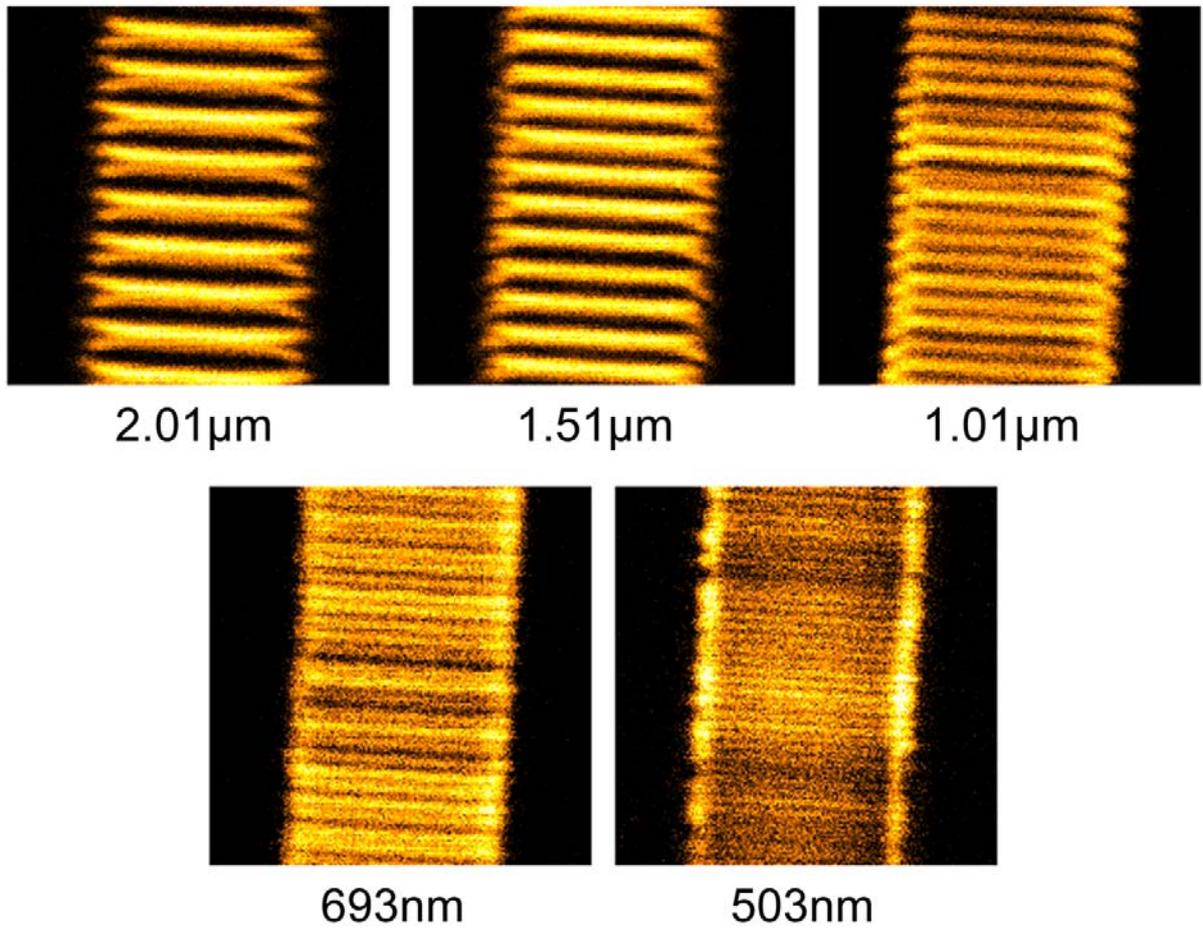

Fig. 2 THG microscope images of WBGs with periods indicated below each pane.

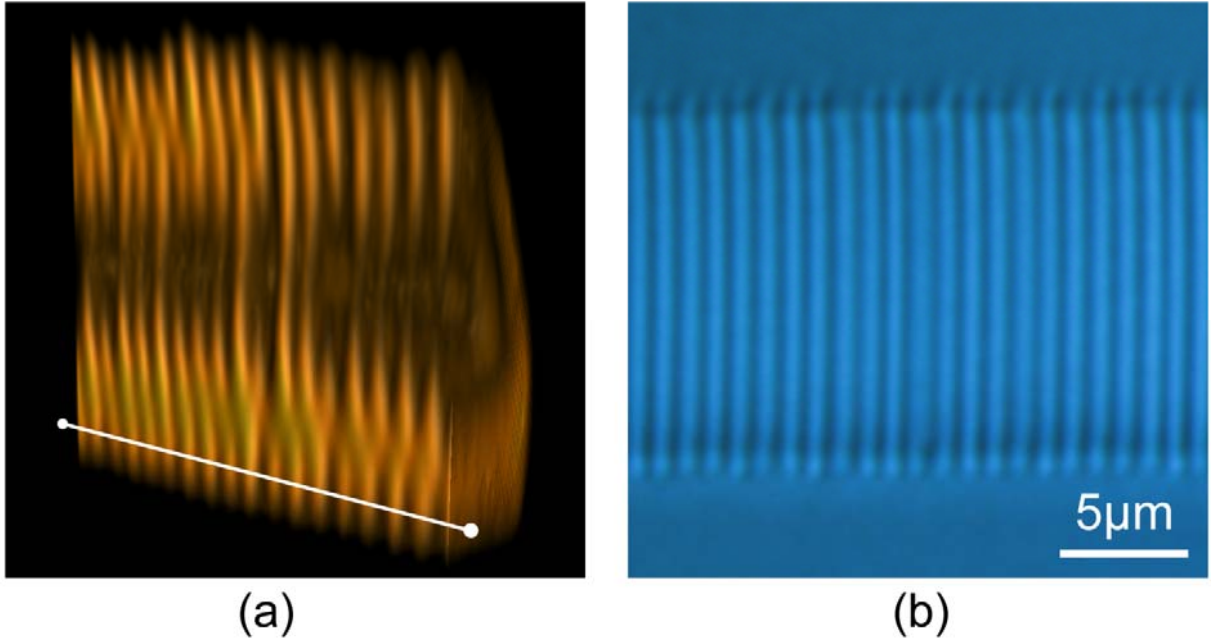

Fig. 3 (a) A still from the animation of a vertically sectioned 1.01 µm period WBG. The writing laser was incident from the top. The white line indicates the position where the TDIC image section (right) is obtained and the full animation is shown in Media 2 (1168 KB). (b) TDIC micrograph of the WBG shown left.